\documentclass[aps,pre,preprint,groupedaddress,floatfix]{revtex4-2}
\usepackage{amsmath}
\usepackage{amsfonts}
\usepackage{mathrsfs}
\usepackage{graphicx}
\usepackage{float}
\usepackage{subfig}
\usepackage{hyperref}
\usepackage{soul}
\usepackage{musicography}

\begin{document}
    \title{Rhythm as an ordered phase of sound: how musical meter emerges in a statistical mechanical model}
    	
    \author{Robert St.Clair}
    \affiliation{Department of Physics, Case Western Reserve University}
    
    \author{Jesse Berezovsky}
    \email{jab298@case.edu}
    \affiliation{Department of Physics, Case Western Reserve University}
    
    \date{\today}

\begin{abstract}
We develop a model of musical rhythm and meter based on optimizing the trade-off between human psychological preferences for perceiving repeated patterns in time with a desire for variety and complexity. By mapping these competing preferences onto analogous quantities in statistical physics, we define an effective free energy which is minimized in the grand canonical ensemble. Using a mean field approximation, we observe phase transitions in the model from disordered events in time to orderings that closely reproduce those seen in music. We then compare the range of rhythmic characteristics predicted by the model to a dataset drawn from compositions by Johann Sebastian Bach, finding generally good quantitative agreement. The results provide a new lens through which to study musical rhythm, and a method for generatively producing rhythms.
\end{abstract}

\maketitle
Despite differences in musical styles across cultures and through time, one of the strongest commonalities among them all is the presence of rhythm -- an ordering of audible events in time. Music theorists characterize rhythms via periodic patterns known as \textit{meters}, specifying time points where events are more or less likely to occur~\cite{Benjamin1984,Longuet1982, lerdahl1996generative, London2012, Sethares2007}. Studies have shed light on perceptual mechanisms that may yield preference for repeated audible events~\cite{Steedman1977, Povel1985, Desain1989, Desain1992, Large1994, Ravignani2018}. Empirical studies have found that innate or enculturated preferences exist across cultures for patterns of sounds that conform to specific meters based on time intervals in integer ratios~\cite{Handel1983,Serafine1989,Desain2003,Keller2005,Large2002,Jacoby2017,Kaplan2022,Jacoby2024}, including in infants as young as seven months~\cite{Hannon2005}. But a question remains: how does a simple preference for repetition lead to the complex and particular meters and rhythms that occur in music?

This work aims to answer the question of how rhythms in music emerge from basic psychoacoustic preferences. We present a model of musical rhythm based on methods from statistical mechanics, drawing from our previous model that captures the emergence of musical harmony~\cite{Berezovsky2019,Din2023, buechele2024renormalization, buechele2024crystals}. In that previous work, the discrete pitches making up a system of musical harmony emerge from the disordered spectrum of non-musical sound via a symmetry-breaking phase transition. The discrete timing of musical events in a rhythm also represents a broken symmetry state as compared to non-rhythmic sequences of audible events. Other work has used concepts from statistics, statistical mechanics, and information theory as top-down, descriptive tools to quantify features of music~\cite{Youngblood1958,Knopoff1983,Hutchinson1987,Colley2019,Levitin2012, Liu2013}. In contrast, our approach is a bottom-up model that aims to show how meter and rhythm can emerge from simple psychological assumptions.

We will show here that a basic model that balances a preference for repeated patterns in time with a desire for variety and complexity does indeed predict a phase transition, where the ordered states closely resemble real musical meters. We map out the phase diagram for the model, and find that the model predicts rhythms with a specific limited range of characteristics. We then compare to a dataset drawn from 42 movements or sections of the six solo Cello Suites by Johann Sebastian Bach (BWV 1007–1012). Specifically we compare the distribution of note lengths in the music to that predicted by the model. We generally observe quantitative agreement with the range of characteristics predicted by the model. Several exceptions are discussed, possibly pointing to the need for models incorporating more detailed psychoacoustic mechanisms or going beyond the mean field assumption used here.

These results provide a new way to study and appreciate rhythms in music, as well as offering a proscription for generatively creating rhythms for algorithmic composition. 

\subsection{Rhythmic Structure}

Rhythm in music refers to an ordered sequence of audible events in time~\cite{Sethares2007, London2012}. Most simply, an audible event could be a click or a strike of a drum. But rhythms can be composed of any combination of audibly perceptible changes that occur at distinct times, such as articulation of a musical tone, or sudden changes in pitch, dynamic, or timbre.

The ordering of events distinguishes a musical rhythm from disordered sound, and even from other non-musical periodic sequences of sounds. Completely disordered sound, such as clicks of a Geiger counter, can be described by a Poisson process, where the probability of an event occurrence is uniform in time. Other sounds, such as a heartbeat, have some periodic structure. Musical rhythm, however, is typically characterized by a \textit{hierarchical meter}~\cite{Longuet1982,Benjamin1984,Palmer1990,lerdahl1996generative,Sethares2007, London2012, Temperley2007}. Meter in music, as in poetry, refers to a pattern of stronger or weaker emphasis. In poetry, the meter refers to a repeating pattern of stressed and unstressed syllables. For example, iambic meter consists of a repeating pattern of syllables (\textit{weak}, \textit{strong}).

\begin{figure}[htb]
\centering
\includegraphics[width=0.5\linewidth]{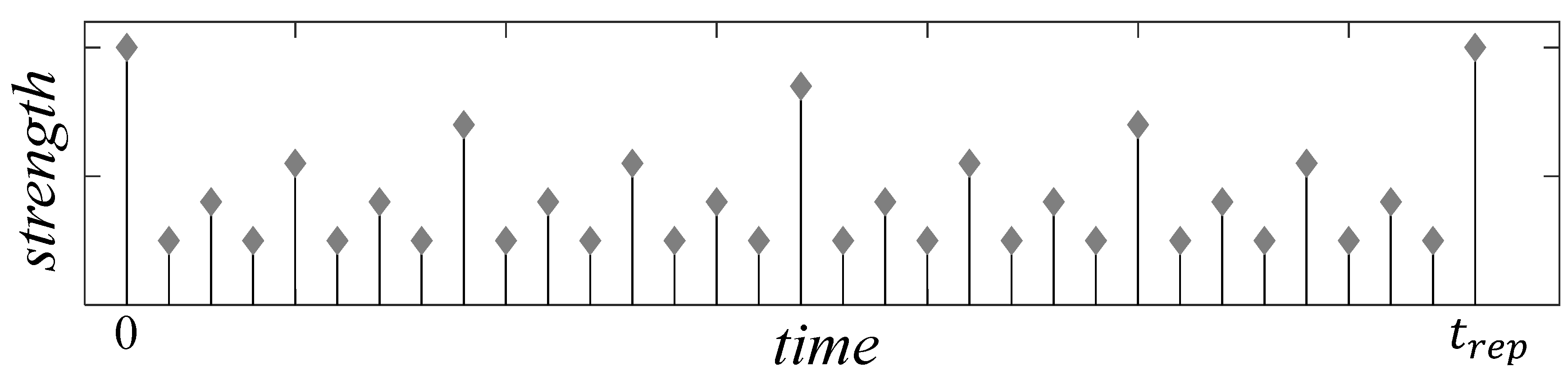}
\caption{\label{fig:meter}Illustration of hierarchical meter found in music.}
\end{figure}

Musical meter is typically a repeating pattern of weaker and stronger emphasis with a hierarchical structure (the ``metric hierarchy''). A common example of a hierarchical pattern repeating with period $t_{rep}$ is shown in Fig.~\ref{fig:meter}. The time point with the strongest emphasis occurs with  period $t_{rep}$. The next strongest time point occurs halfway between the strongest points, dividing the period into two intervals of $t_{rep}/2$. These two shorter intervals are again divided in two by the third strongest time points. There are two time points at the third-strongest level. This process of division can be repeated indefinitely (though eventually, the time divisions become imperceptibly close together.) The metric structure shown in Fig.~\ref{fig:meter} is referred to as ``common time'' and is notated musically as a ``4/4 time signature.'' This is the most common meter in Western music. Other meters, such as 3/4 or 6/8 time can be understood in the same way, but where a single level of the hierarchy is defined by a division by 3 instead of by two. 

The meter translates into aspects of music in several ways. Stronger and weaker emphasis can simply manifest as louder or quieter (accented or unaccented) audible events, or can govern the placement of pitches that have stronger or weaker roles in the melody and harmony. But the meter also affects the ordering of the events~\cite{lerdahl1996generative}. For example, an event is more likely to occur at a time point with stronger emphasis~\cite{Temperley2007}. Similarly, the time between events is likely to be longer following an event at a stronger time point~\cite{Temperley1999, Palmer1990, Longuet1982}. 

For simplicity here, we will refer henceforth to an ``event'' as the onset of a ``note'' that then continues until the next event. The span between events (often called the inter-onset interval) is then referred to here as the note length. In this terminology, notes are more likely to begin at stronger time points, and the stronger the point, the more likely that a longer note occurs there.

The metric hierarchy describes the lowest level of temporal organization in music. As described in Ref.~\cite{lerdahl1996generative}, other patterns of repetition are found at longer time scales. The meter describes patterns at the level of individual beats. Then patterns of several beats, referred to as rhythmic ``motives'' might be repeated in some temporal arrangement. Still larger phrases or sections may be repeated exactly or with some variation. Though this larger scale ``grouping'' structure can also be described hierarchically, we only expect the lower level metric structure to obey the quantitative statistics we focus on here.

Questions that we aim to address here include, why musical rhythm is based on such a hierarchical meter, how many levels of the hierarchy are needed in practice, and how the meter is quantitatively related to the probabilities of note onsets and note lengths.

\section{Model}

The model we present here is based on the balance between a psychoacoustic preference for the perception of repeated patterns of events in time, with a desire for variety and complexity in music. This follows the approach used previously to model harmony in music, where the psychoacoustic preference is for the perception of consonant combinations of pitches. A pattern of events that maximizes repetition would be a completely regular repeating series. This represents a single possible rhythm (up to a rescaling by the time interval between the events). In order to have music with a variety of rhythms allowing for a wider range of expression, the desire to maximize repeated patterns must be balanced by a desire to maximize the number of possible rhythms. These factors can be mapped onto analogous quantities from thermodynamics. The desire to maximize repeated patterns (``rhythmicity'') is analogous to a system that tends to minimize its energy, and the desire to maximize the (log of) the number of possible rhythms is analogous to entropy $S$. This motivates our general strategy of borrowing techniques from statistical mechanics to study the orderings that occur in musical rhythm.

The system we consider is a set of $\mathcal{N}$ ``note onsets'' that occur at times $\{t_i\}^\mathcal{N}$. A note onset refers generally to an audible event. Given a method for calculating the total rhythmicity $R_{tot}$ of a set $\{t_i\}^\mathcal{N}$, we will seek to minimize a free energy $F = -R_{tot} - TS - \mu \mathcal{N}$. The temperature $T$ controls the desired trade-off between rhythmicity and variety, and the chemical potential $\mu$ controls the concentration of notes in time. Both $T$ and $\mu$ are in the same units as rhythmicity $R$ (and thus the constant analogous to the Boltzmann constant $k_B=1$). These units are arbitrary, and will be unlabeled below, but values of $R$, $\mu$, and $T$ can be meaningfully compared to each other.

For simplicity, we will discretize time into a set of bins centered at equally spaced times $\tau_j$. The duration of these bins could be set to be the minimum time that the human auditory system can distinguish separate events or that a sequence of events can be parsed as a rhythm ($\sim 10$~ms and $\sim 100$~ms, respectively)~\cite{London2012}, or could be set to a larger interval for convenience of calculation. The discretization of time also accounts for the flexibility in timing of events in actual musical performance~\cite{Jacoby2024}. If a note onset time $t_i$ falls within time bin $j$, then we define the occupancy $B_j = 1$, and if not $B_j=0$. We will assume that at most one note onset can occur in a given time bin. In this way, the set of onset times $\{t_i\}^\mathcal{N}$ is replaced by a set of occupancies $\{B_j \in \{0,1\} \}$ with $\sum B_j = \mathcal{N}$. We will formulate a simple model for $R_{tot}$ in terms of $B_j$ and study the statistics of the $B_j$.

\subsection{Rhythmic Perception}

It has often been proposed that the human perception of rhythm is based on expectation: the spacing between an event at time $t_1$ and another event at time $t_2$ is used to formulate expectations about the occurrence of another event at $t_3$~\cite{Longuet1982, Huron2008,Desain2003,Ravignani2018,Jacoby2024,Kaplan2022}. In the simplest case, two events create an expectancy for a third event with equal time spacing. Some models also include a preference for other small integer ratios of the spacings. Experimental studies have attempted to identify rhythmic ``priors,'' simple rhythmic motifs that listeners use to perceive and classify rhythms~\cite{Desain2003, Jacoby2017, Kaplan2022, Jacoby2024}. These studies have confirmed the universal preference for time intervals in small integer ratios, but also found that listeners from different cultures show preferences weighted towards different patterns of ratios, demonstrating the role of enculturation in perception of rhythm and meter. Many models of meter perception are based on the idea that the spacings of rhythmic events in time entrain internal clocks that gives rise to rhythmic expectancies~\cite{Povel1985, Desain1989,Desain1992, Large1994,Large2002,London2012}. 

Based on these ideas of rhythmic perception, we construct a minimal model: We posit that three events are perceived as rhythmic if they are equally spaced. If events occur in time bins $i$, $j$, and $k$, ($i<j<k$) then they are perceived as rhythmic if $k-j = j-i$. We can quantify the perception of rhythm arising from these three bins as $R_{ijk} = R_0 B_iB_jB_k \delta_{k-2j+i}$, where $\delta$ is the Kronecker delta.  This model corresponds to the assumption of a preference only for 1:1 ratios of time intervals. However, there is no restriction that the three equally-spaced events need to be adjacent, a fact that leads to an entrainment-like effect where a regularly repeating sequence of events leads to a self-reinforcing proliferation of equally spaced triples.

One would expect the constant of proportionality $R_0$ to at least depend on the time difference $\tau_k-\tau_j$, or possibly on other factors. Certainly if $\tau_k - \tau_j$ becomes too large, $R_0$ should go to zero as the separation between beats becomes too long for the perception of a pattern. It is also possible that $R_0$ would decline at smaller $\tau_k-\tau_j$ as patterns become harder to discern. Studies have shown that perception of a regular beat is difficult outside the range of $\sim 100-2000$~ms spacing~\cite{London2012}. The limit of $\tau_k-\tau_j \rightarrow 0$, however, is already precluded by the discretization of time. $R_0$ might also depend on the overall fraction of occupied bins, where very dense (fast) rhythms might have a lower $R_0$ because the high rate of equally spaced triples might start to make each individual triple harder to discern. Below, we will make some simplifying and reasonable assumptions for the behavior of $R_0$.

The total perception of rhythm $R_{tot}$ must then be calculated by combining the $R_{ijk}$. This combination might not be simply additive and might depend on various factors. Here we will set $R_0 = 1/(N\tilde{B}^2)$ for $k-j \leq N$ and $R_0=0$ for $k-j>N$, where $N$ is a constant specifying a hard cut-off time span beyond which rhythmic patterns are no longer perceived, and $\tilde{B}$ is the mean occupancy of the time bins. This choice of prefactor normalizes a regular repeating rhythm with any period less than $N$ to have rhythmicity per site $\tilde{R}_{tot}=1$ (see Methods section). We can then sum over all $R_{ijk}$ as

\begin{equation}
\label{eq:Rtot}
R_{tot} = \frac{1}{N \tilde{B}^2} \sum_{j = -\infty}^\infty B_j \sum_{s=1}^N B_{j-s}B_{j-2s}.
\end{equation}

 When a note occurs in bin $j$, a nonzero term is added to the outer sum if and only if there were notes present a time $s$ and $2s$ bins previously. We can additionally calculate the \emph{change} in rhythmicity for a given site $k$ if $B_k$ were switched from 0 to 1, removing the need for an infinite sum. Since $B_k$ can be present in Eq.~\ref{eq:Rtot} as $B_j$, $B_{j-s}$, or $B_{j-2s}$, we can substitute in the values $k = j$, $k = j - s$, and $k = j - 2s$ to this equation and solve for $\Delta R_k$ for these three cases:

\begin{equation}
\label{eq:dR}
\Delta R_k = \frac{1}{N \tilde{B}^2} \sum_{s = 1}^N (B_{k-s}B_{k+s}+B_{k+s}B_{k+2s}+B_{k-s}B_{k-2s}).
\end{equation}

\subsection{Mean field approximation}

Given the expression for $\Delta R_k$ (Eq.~\ref{eq:dR}), we can study the equilibrium occupancies $\{B_j\}$ that minimize the free energy at some $\mu$ and $T$. In the grand canonical ensemble, the mean occupancy of site $k$

\begin{equation}
    p_k = \langle B_k \rangle =  \frac{1}{1+\exp{\left[ -(\Delta R_k + \mu)/T \right]}}.
\end{equation}

In this equation, $p_k$ depends on many other $B_j$, through $\Delta R_k$. To study this system, we make some additional simplifying assumptions. We will first assume that the average occupancy $p_k$ of each site can be described by a probability independent of the particular occupancy $B_j$ of the other bins, but only dependent on the average occupancies $p_j$ of the other bins. This is the standard mean field approximation, which can also be thought of as extending the range of the interaction $N$ to infinity.  We will also assume that the mean occupancies $p_l$ are periodic with period $L$.  

With these assumptions, we have $\tilde{B} = \frac{1}{L} \sum_{l=0}^{L-1} p_l$, and 

\begin{equation}
\label{eq:dR with p}
\Delta R_k = \frac{L}{(\sum p_l)^2} \sum_{l=0}^{L-1}(p_{k-l} p_{k+l} + 2p_{k+l}p_{k+2l}).
\end{equation}

Finally, we have a self-consistent set of equations for the $L$ unique values of $p_l$:

\begin{equation}
\label{eq:p}
p_l = \frac{1}{1+\exp{\left[ -(\Delta R_l + \mu)/T \right]}}.
\end{equation}

\section{Results}

\subsection{Two-Site Model}

We will start with the simplest case of $L=2$ and see that even in this simple case, realistic rhythmic ordering occurs as a phase transition from Poissonian random events.

 For $L=2$ it is illustrative to reparameterize $p_0$ and $p_1$ as a ``magnetization'' per bin $m = (p_1 - p_0)/2$ and the average density $n = (p_0 + p_1)/2$. Here, $m$ is an order parameter representing the difference between levels of the metric hierarchy (with at most two levels for $L=2$.) We will also add a forcing term to $\Delta R$, $-h(-1)^l$ at site $l$, where $h$ is a parameter analogous to magnetic field. $h$ will typically be set to zero, but allows us to calculate the Landau free energy. Substituting $m$ and $n$, Eq.~\ref{eq:dR with p} can be rewritten as

\begin{equation}
\label{eq:R_l m and n}
\Delta R_l = 3 + \left(\frac{m}{n}\right)^2 - (-1)^l\left(h + 2\frac{m}{n}\right).
\end{equation}

We can then plug this into Eq.~\ref{eq:p} for $l=0$ and $l=1$, again rewriting to obtain two equations with $n$ and $m$ on both sides. Adding and subtracting these two equations we obtain, with some rearrangement:

\begin{equation}
-T\log{\left[ \left(\frac{1-(n-m)}{n-m}\right) \left(\frac{1-(n+m)}{n+m}\right) \right]} = 6 + 2\left( \frac{m}{n} \right)^2 + 2\mu
\label{eq:nmsum}
\end{equation}

and 

\begin{equation}
T\log{\left[ \frac{1-(n-m)}{n-m}\frac{n+m}{1-(n+m)} \right]} = 2\left( h+ 2\frac{m}{n} \right).
\label{eq:nmdiff}
\end{equation}

\begin{figure}
\centering
\includegraphics[width=0.5\linewidth]{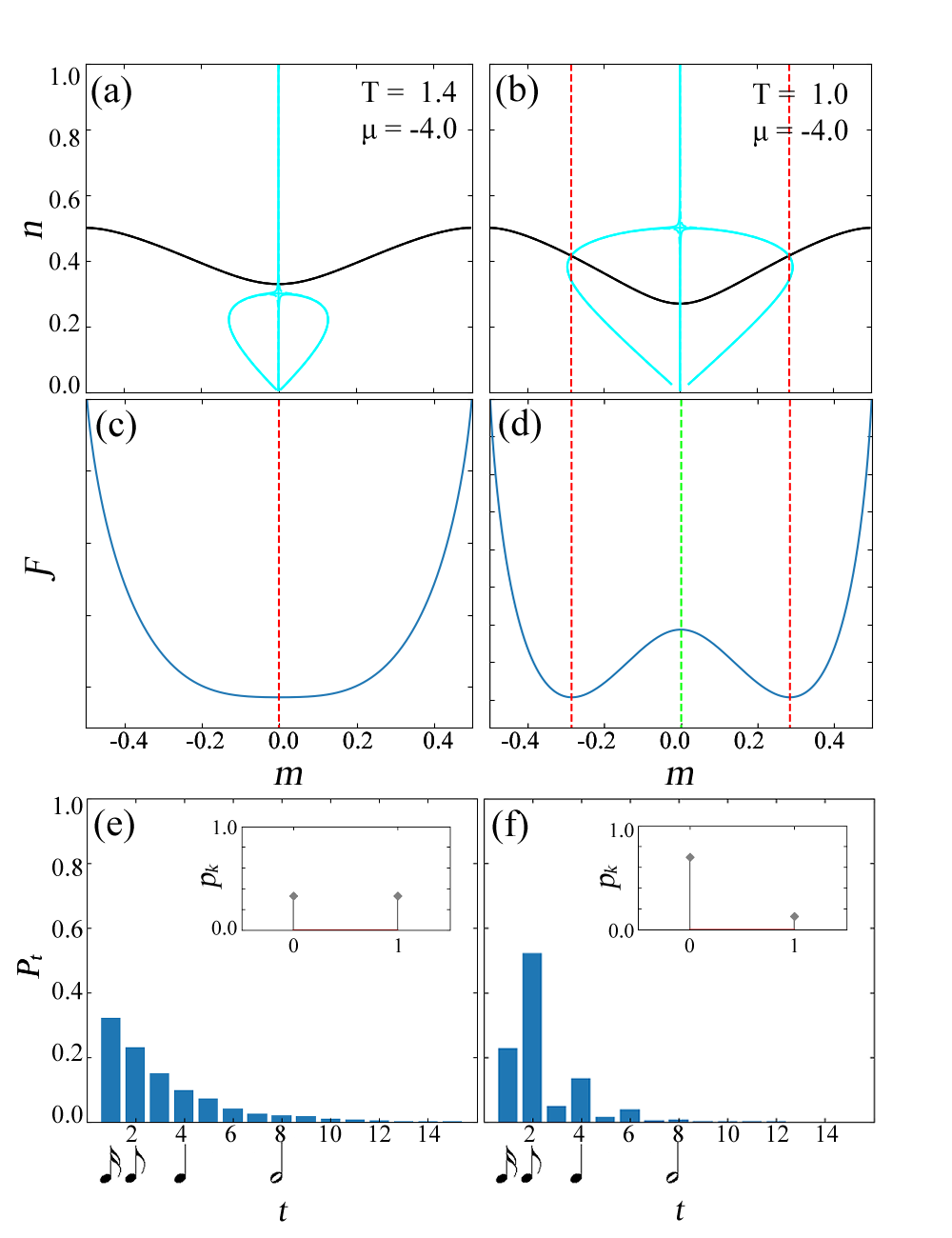}
\caption{\label{fig:twosol}The two-site ($L=2$) model. (a) and (b) show solutions of the system of two equations for $m$ and $n$ at intersections between the black and cyan curves. One solution is seen at higher $T$ in (a), and three solutions at lower $T$ in (b). (c) and (d) show the Landau free energy $\mathcal{F}$ in arbitrary units vs. order parameter $m$, corresponding to the parameters given in (a) and (b), respectively. Minima (maxima) of $\mathcal{F}$ are indicated by red (green) dashed lines. (e) and (f) show probabilities $P_t$ of note lengths $t$, corresponding to parameters in (a) and (b), respectively. Insets show the values of $p_k$ obtained in the two cases from a stable solution of $m$ and $n$.}
\end{figure}

As shown in Fig.~\ref{fig:twosol}(a) and (b), plotting Eqs.~\ref{eq:nmsum} and \ref{eq:nmdiff} with $h=0$ and some values of $T$ and $\mu$ graphically illustrates the self-consistent solutions at the intersections of the two curves. Fig.~\ref{fig:twosol}(a) shows Eqs.~\ref{eq:nmsum} (black) and \ref{eq:nmdiff} (cyan) at $T=1.4$ and $\mu=-4$. A single solution is seen at $m=0$ and $n\approx 0.4$. Lowering the temperature to $T=1.0$, Fig.~\ref{fig:twosol}(b) shows how the two curves have shifted, now with three intersections, representing three solutions, at $m=0$ and $m \approx \pm 0.3$.

To assess the stability of the solutions seen graphically in Fig.~\ref{fig:twosol} (a) and (b), we calculate the Landau free energy, $\mathcal{F}(m)$, by using the thermodynamic identity 
\begin{equation}
\label{eq:dF/dm}
\left. \frac{\partial F}{\partial m} \right\rvert_{\mu,T} = h
\end{equation}
as described in the Methods section below.

The Landau free energy $\mathcal{F}(m)$ is shown in Fig.~\ref{fig:twosol}(c) and (d), for the parameters in (a) and (b) respectively. At the higher temperature (Fig.~\ref{fig:twosol}(c)) the single solution is seen to be a minimum of $\mathcal{F}$, and therefore is stable. At lower temperature (Fig.~\ref{fig:twosol}(d)), the solution at $m=0$ is now a maximum of $\mathcal{F}$, and the two solutions at $m \neq 0$ are stable minima. This is the prototypical behavior of a second order phase transition where the stable solution at zero order parameter becomes unstable and bifurcates into two stable minima with nonzero order parameter. The system then chooses one of the two minima randomly in the process of spontaneous symmetry breaking. 

The characteristics of the disordered ($m=0$) and ordered ($m\neq0$) phases shown in Fig.~\ref{fig:twosol} can be further investigated by calculating the probabilities $P_{t}$ of different note lengths, where the index $t=1,2,3...$ specifies the number of time bins spanned between note onsets. As discussed above, $p_k$ is the probability of a note onset in time bin $k$. We can then combinatorially calculate the probability of different note lengths as the time between onsets. For $L=2$, it is straightforward to calculate the $P_t$. If a note onset occurs, there is probability $p_0/(p_0+p_1)$ that it is in an even bin and probability $p_1/(p_0+p_1)$ that is in an odd bin. Then the probability that the next bin also contains an onset yields $P_1 = 2 p_0 p_1 / (p_0+p_1)$, given by summing the cases where the initial onset is in an odd or even bin. $P_2 = \left[ p_0^2 (1-p_1) + p_1^2(1-p_0)\right]/(p_0+p_1)$, given by the probability that note onsets do not occur in adjacent bins, but in the next bin instead. This calculation can be iterated to find any $P_t$. Alternatively, the $P_t$ can be calculated numerically by generating a sufficiently long sequence of pseudorandom numbers $\{r_k\}\in[0,1)$, choosing note onsets at bins $\{ (p_k > r_k) \}$ and building a histogram of inter-onset intervals. We compare to the analytical results to ensure that the sequence is sufficiently long.  

Note length probabilities $P_t$ are shown in Fig.~\ref{fig:twosol}(e) and (f) for the disordered and ordered cases respectively. In the disordered case ($m=0$), the inset to Fig.~\ref{fig:twosol}(e) shows the two equal values $p_0$ and $p_1$. Because the probability of an event is the same in every time bin, the resulting statistics are Poissonian, with an exponential distribution of note lengths. On the horizontal axis, we compare to note lengths typically found in music. Here, we arbitrarily assign the shortest note length to be a sixteenth note $(\musSixteenth)$. Other common note lengths are related to this by successive factors of two: eighth notes $(\musEighth)$, quarter notes $(\musQuarter)$, half notes $(\musHalf)$ and whole notes $(\musWhole)$. We see that in the disordered phase there is no preference in the values of $P_t$ for these commonly used note lengths.

In the ordered phase, the $P_t$ are qualitatively different than in the disordered case, as seen in Fig.~\ref{fig:twosol}(f). Here, $p_0 \neq p_1$, as shown in the inset, and the $P_t$ do not follow an exponential distribution. Instead, we see that some note lengths are increased in probability, and others are suppressed. $P_2$ is now the maximum, shown here as an eighth note. The next most probable note lengths are $P_1$ (sixteenth note) and $P_4$ (quarter note), at half and double the most common length, respectively. Next most probable are $P_3$ and $P_6$ which are musically notated as ``dotted'' eighth notes $(\musEighthDotted)$ and dotted quarter notes $(\musQuarterDotted)$. The lower probability $P_5$ does not correspond to any note symbol in common musical notation, suggesting that this length should be relatively less probable. We also see that longer common note lengths ($P_8$ and $P_{16}$, half and whole notes) are rather low probability. This constitutes a prediction for the quantitative relationship between the metric hierarchy and note onsets/lengths. Below, we will compare results of the model to rhythms observed in music.

We next study the same model, but now with a larger number ($L=8$) of independent occupancy probabilities.

\subsection{Eight-site model}

As we increase $L$, we find similar behavior to the $L=2$ case, but with some additional subtleties. For $L>2$, the graphically-aided method used above is not practical for solving the system of $L$ equations. Instead, we use an iterative method that begins from $L$ initial values $p_l^0$ and converges to a set of $p_l$ that solve Eq.~\ref{eq:p} (see the Methods section for more detail.) 

\begin{figure*}[h!]
\centering
\includegraphics[width=1.0\textwidth]{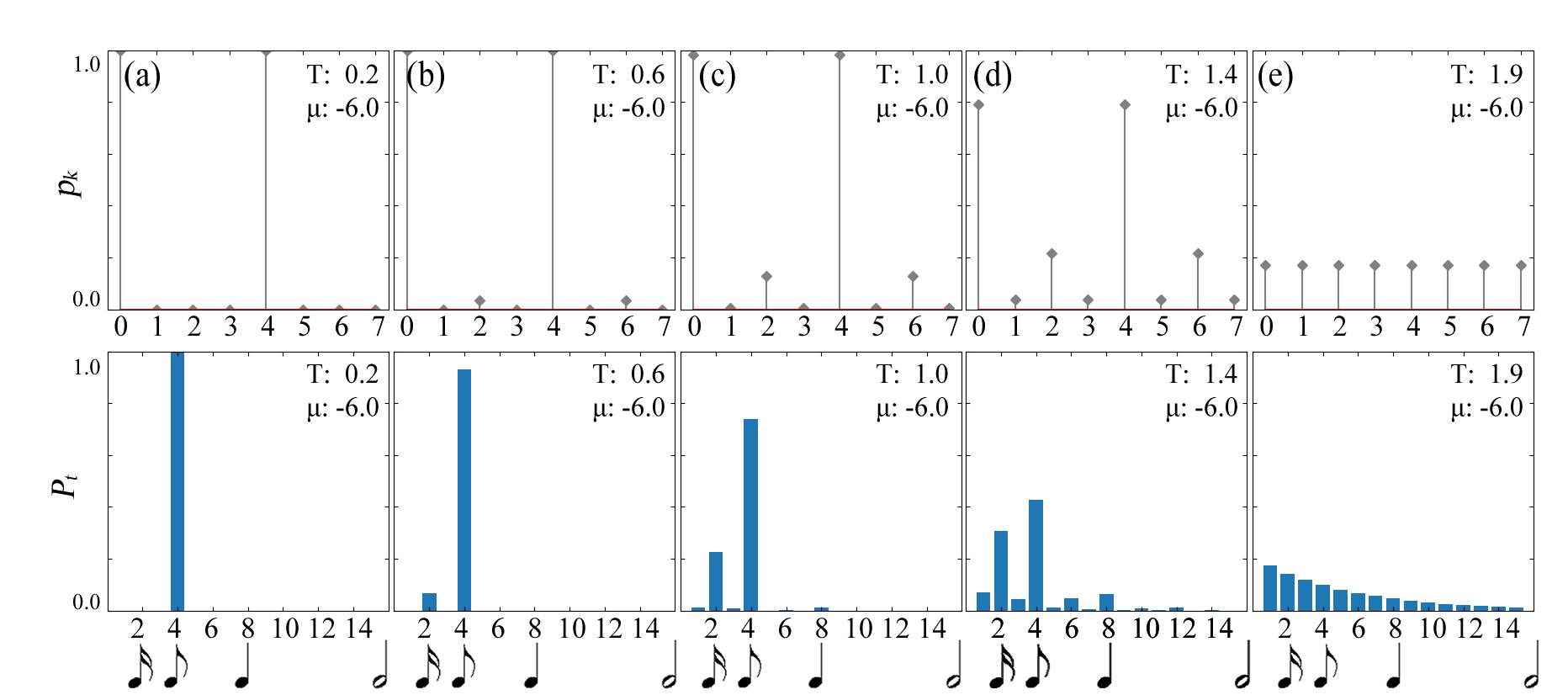}
\caption{\label{fig:tsweep} (a)-(e) show results of the eight-site ($L=8$) model, at increasing $T$ and fixed $\mu=-6.0$. The top row shows mean occupancies $p_k$ of each independent site $k$. The bottom row shows probabilities $P_t$ of note lengths $t$.}
\end{figure*}

Figure~\ref{fig:tsweep} shows representative results from the $L=8$ model, with temperature $T$ increasing from left to right, and a constant $\mu = -6.0$. The calculation is initialized with uniform random $p_l^0$, and the resulting $p_l$ are cyclically shifted to place the largest $p_l$ at $l=0$, for easier comparison. The supplementary audio files, generated by sampling from the probabilities $p_l$ (see Methods), demonstrate the results of Fig.~\ref{fig:tsweep}(b)-(e).

At the lowest temperatures (Fig.~\ref{fig:tsweep}(a)), we see a nearly completely regular series of events with $p_l \approx 1$ at $\tau_0$, $\tau_4$, and then due to periodic boundary conditions repeating regularly thereafter. By construction, a regularly repeating pattern with any period has $\tilde{R}_{tot} = 1$. The 4-bin period seen here is therefore set by the value of $\mu$. Note that here additional periodicity has emerged in addition to the assumed $L$-site periodicity, showing that periodic patterns of $p_l$ are not just put into the model by hand. This repeated pattern of note onsets results in overwhelmingly a single note length, with $P_4 \approx 1$, and all other $P_t \approx 0$, as shown at the bottom of Fig.~\ref{fig:tsweep}(a). We arbitrarily assign this note length to be an eighth note.

As the temperature is increased to $T=0.6$ in Fig.~\ref{fig:tsweep}(b), we see that $p_0$ and $p_4$ are still very close to unity, but now $p_2$ and $p_6$ are visibly nonzero. As such, the note lengths are now still dominated by eighth notes at $P_4$, but those time spans are occasionally divided by a note onset at $\tau_2$ or $\tau_6$, yielding some noticeable value of $P_2$, corresponding to sixteenth notes (see Supplementary Audio File 1).

At $T=1.0$ (Fig.~\ref{fig:tsweep}(c)), $p_0$ and $p_4$ are now visibly slightly less than unity, and $p_1$, $p_3$, $p_5$, and $p_7$ are now very slightly visibly nonzero. The note length probabilities are still dominated by eighth notes. But now the decreased $p_0$ and $p_4$ yield some probability of longer notes, primarily quarter notes at $P_8$, and the increased $p_l$ at odd $l$ yield some small probability of $32^\mathrm{nd}$ notes at $P_1$ (see Supplementary Audio File 2). Note the similarity of the even note length probabilities here to those in the $L=2$ model shown in Fig.~\ref{fig:twosol}(f). Here again, the dotted notes at $P_3$ and $P_6$ have some small probability, but the still-less-common note lengths at $P_5$ and $P_7$ are not visibly present.  

Increasing the temperature further to $T=1.4$ (Fig.~\ref{fig:tsweep}(d)) continues the trends seen in (a)-(c). Now the metric hierarchy is clearly visible in the values of $p_l$, as illustrated in Fig.~\ref{fig:meter}. With all probabilities significantly deviating from zero or one, a wider range of note lengths are probable, though still with the same ordering of likelihood for the different $P_t$ (see Supplementary Audio File 3).

As the temperature is increased in Figs.~\ref{fig:tsweep}(a)-(d), smooth evolution of the $p_l$ is observed, maintaining the same 4-site translational symmetry. This situation is now changed, however, in Fig.~\ref{fig:tsweep}(e) ($T=1.9$), where the $p_l$ have translational symmetry by a single bin. This higher-symmetry state is the disordered phase, which again shows an exponential distribution of note lengths, with no preference for musically common notes (see Supplementary Audio File 4).

\subsubsection{Phase diagram for the 8-site model}

\begin{figure}[h!]
\centering
\includegraphics[width=0.75\textwidth]{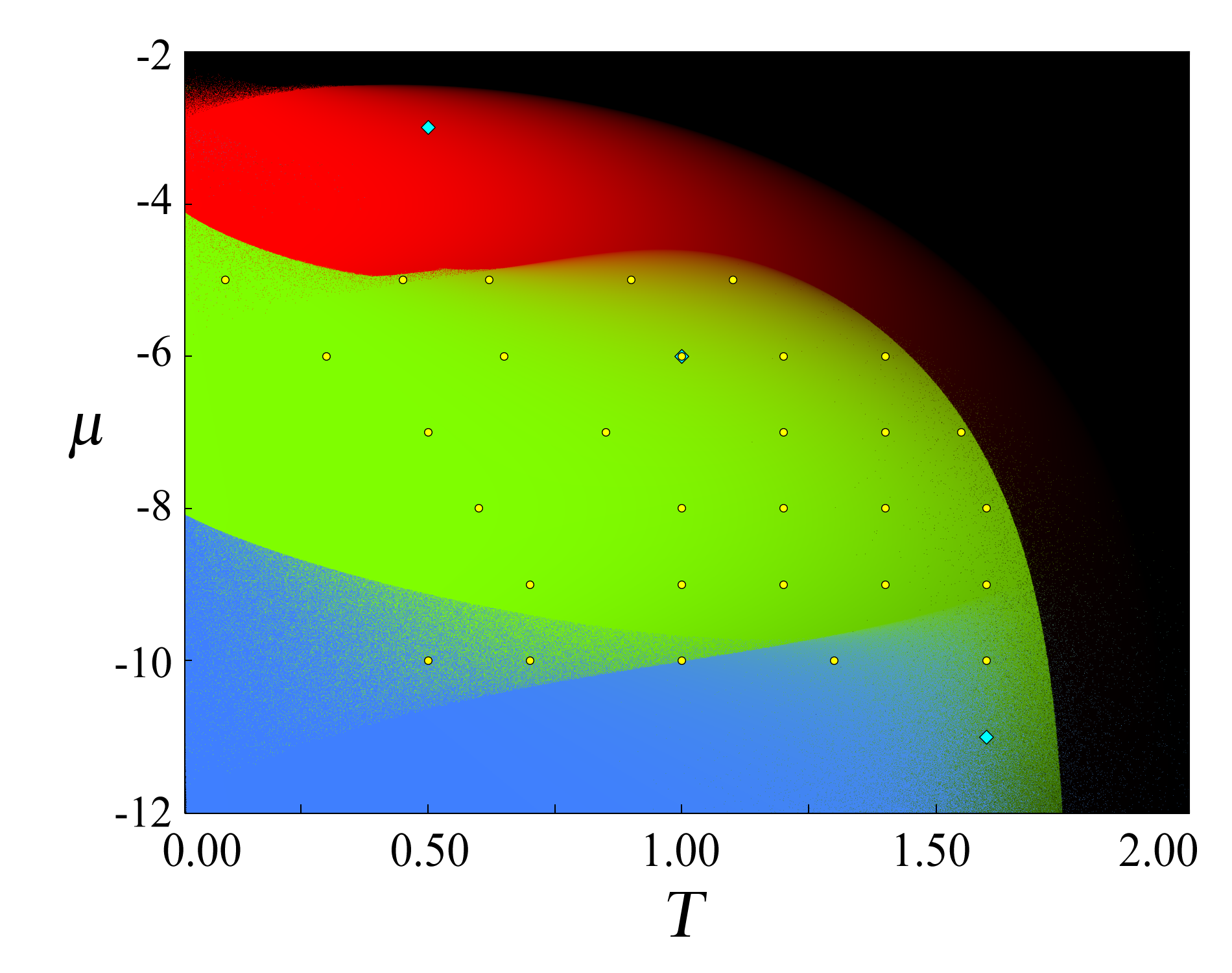}
\caption{\label{fig:phase}Phase diagram of 8-site ($L=8$) model. Random initializations were used to calculate the mean occupancies $p_k$ of each site, yielding order parameters $m_2$, $m_4$, and $m_8$. Each $(T,\mu)$ point is colored by RGB triplet $(|m_2|, |m_4|, |m_8|)$. Points of interest at values of $T$ and $\mu$ used in plots are marked by yellow circles (example note length probabilities, Fig.~\ref{fig:dists}) and cyan diamonds (examples of self-similar states, Fig.~\ref{fig:degen}).}
\end{figure}

We explore the results of the 8-site model by constructing a phase diagram as a function of $\mu$ and $T$. Over a grid of values spanning ranges in $\mu$ and $T$, we calculate $p_l$ using the iterative method described above, starting from uniform random initial conditions. The resulting $p_l$ are found to be symmetric by translation of 1, 2, 4, or 8 sites, and have the hierarchical structure as typified by Fig.~\ref{fig:meter}. Following the $L=2$ case, where the order parameter was $m=p_1-p_0$, we can generalize to multiple order parameters 
\begin{equation}
m_q = \frac{q}{L} \sum_{l=0}^{L-1}p_l e^{2 \pi i l/q},
\end{equation}
where $q>1$ is an integer that divides $L$, yielding the multiple levels of the hierarchy. The $m_q$ are defined such that $0 \leq |m_q| \leq 1$, with $|m_q| =1$ representing regular repetition of notes with length $q$. For $L=8$, we then have three order parameters $m_2$, $m_4$, and $m_8$, describing the two-site, four-site, and 8-site periodic structure. The amplitudes $|m_q|$ combine to yield the differences in $p_k$ between levels of the hierarchy, and the complex phase reflects the spontaneously broken symmetry of an arbitrary cyclic permutation of the $p_k$. We can color code the order parameters as an RGB triplet $(|m_2|, |m_4|, |m_8|)$. The resulting phase diagram is shown in Fig.~\ref{fig:phase}. Four distinct colors are seen (black, red, green, and blue) corresponding to four phases described below. Some phase boundaries show second-order behavior where the color changes in a continuous way such as the red-green boundary around $T=1$, $\mu=-4.5$. Other regions appear speckled, indicating multiple stable solutions in different phases suggestive of a first-order transition. The random speckle occurs because of different random initial conditions leading to different stable solutions.

The four distinct colors in the phase diagram in Fig.~\ref{fig:phase} are to be understood as follows. Black corresponds to the RGB values $(0,0,0)$, meaning that all $p_l$ are equal, and hence all $m_q = 0$. This is the disordered phase shown in Fig.~\ref{fig:tsweep}(e). The red color corresponds to RGB values $(|m_2|, 0, 0)$. This corresponds to a phase with two-site translational symmetry, essentially the same as the phase explored in the two-site model in Fig.~\ref{fig:twosol}. The green region then corresponds to RGB values $(|m_2|, |m_4|, 0)$. This is the phase with four-site translational symmetry, as shown with $\mu=-6$ in Fig.~\ref{fig:tsweep}(a)-(d). Finally, the blue region corresponds to all three order parameters nonzero, corresponding to the phase with only the assumed 8-site periodicity.  

\subsubsection{Self-similarity}

By construction, the total rhythmicity per site $\tilde{R}_{tot}=1$ for any regularly repeating sequence of events. Such a regularly repeating pattern is the ground state of this model, approached as $T\rightarrow 0$. This ground state appears at the left side of the phase diagram in Fig.~\ref{fig:phase}. As $\mu$ becomes more negative, the regular repetition of the ground state is maintained, but with the periodicity changing. Lower $\mu$ favors an equilibrium state with a lower concentration of events. The ground state in the black region corresponds to all $p_l \rightarrow 1$, whereas the ground state in the blue region corresponds to every 8th $p_l \rightarrow 1$, and all other $p_l \rightarrow 0$. The wedge-shaped regions of bistability seen at low $T$ correspond to regions where the repeating pattern has some periodic $p_l \approx 1$ and the other $p_l \approx 0$. 

\begin{figure}
\centering
\includegraphics[width=0.9\textwidth]{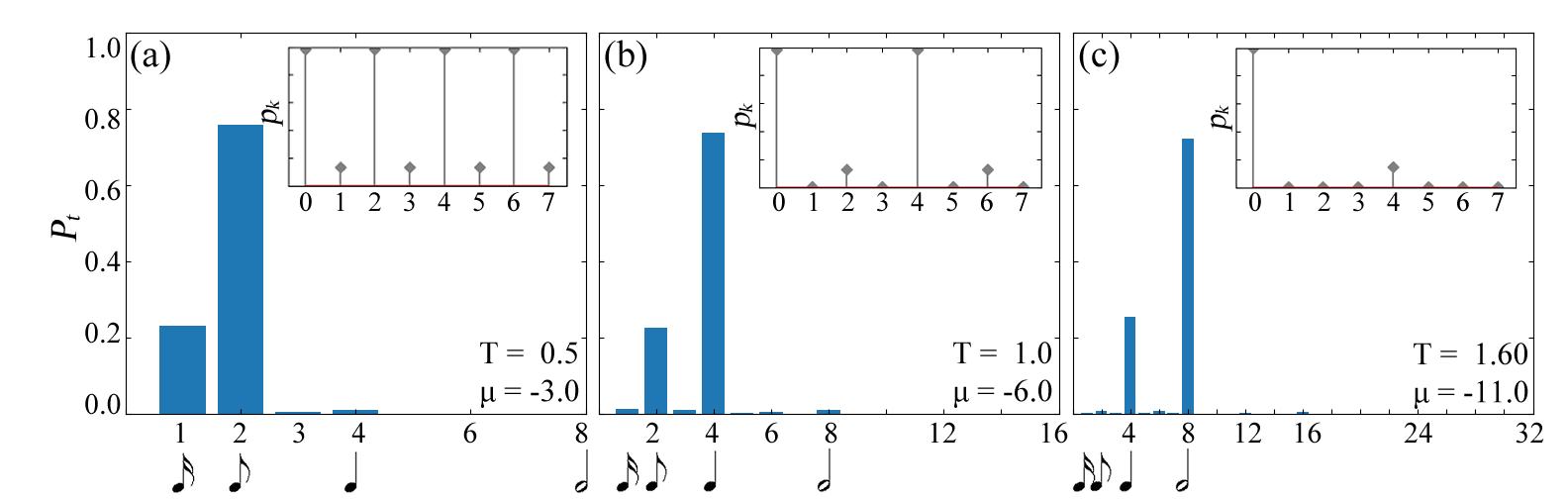}
\caption{\label{fig:degen}Probabilities $P_t$ of note lengths $t$ at the three given $(T,\mu)$ values, also indicated by the cyan diamonds in Fig.~\ref{fig:phase}. Time bin mean occupancies $p_k$ are shown in the insets.}
\end{figure}

The self-similarity across timescales does not only occur near the ground state, but also at higher temperatures. Fig.~\ref{fig:degen}(a), (b), and (c) show results from three $(T,\mu)$ points in the red, green, and blue regions, respectively. These points are indicated by cyan diamonds in Fig.~\ref{fig:phase}. (The solution in Fig.~\ref{fig:degen}(c) corresponds to the blue color in the phase diagram.) The values of $p_k$ shown in the insets are quite similar, but with the time scale altered by factors of two. The resulting note length distributions are also quite similar, up to a doubling of note lengths. In each case, a single time bin is assigned to be a sixteenth note, and the horizontal axes are rescaled by factors of two to highlight the self-similarity. Referring to the phase diagram in Fig.~\ref{fig:degen}, we can see that these self-similar points occur at higher $T$ as $\mu$ is increased. This is also reflected by the fact that the bistable wedges at lower $T$ grow as $\mu$ is decreased.

The self-similarity across timescales supports the freedom to discretize time into bins of arbitrary size. Whether the time bins are set to be the shortest perceptible separation of events, or if the bins are set to some shortest note length of interest, similar event probabilities vs. time should occur in both cases by adjusting $\mu$ appropriately.

\subsection{Mapping out the space of rhythms}

We have seen that this model reproduces familiar hierarchical metric structures of musical rhythm. But looking closer, we will see that not just any hierarchical event probabilities are possible. For example, the hierarchical structure illustrated in Fig.~\ref{fig:meter}, with six significant levels of the hierarchy, does \textit{not} emerge in this model. We will now map out the types of rhythms that do emerge. Because of the self-similarity described above, we will restrict our attention to one region of the phase diagram in Fig.~\ref{fig:phase}.

\begin{figure*}
\centering
\includegraphics[width=1\textwidth]{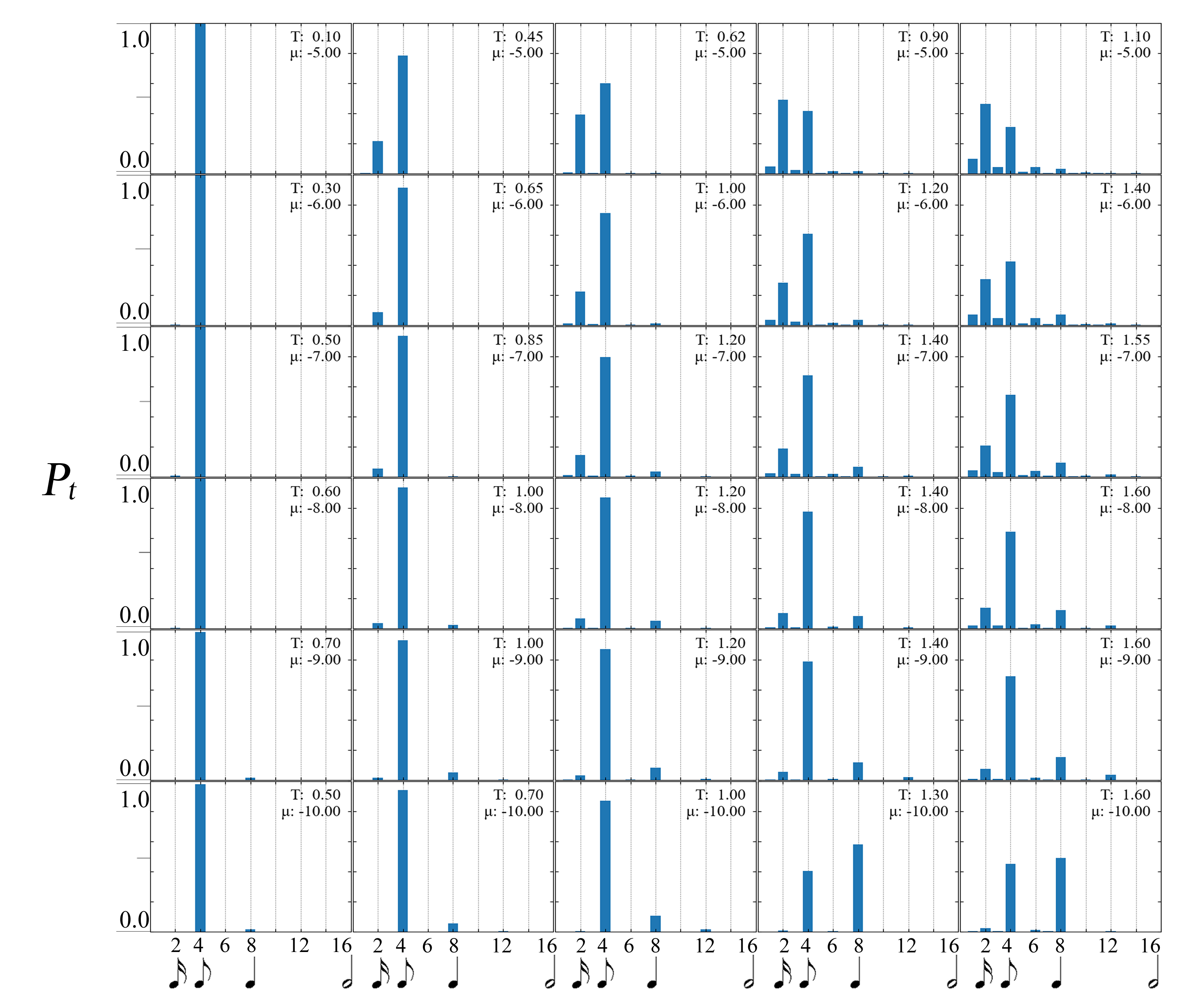}
\caption{\label{fig:dists}Probabilities $P_t$ of note lengths $t$ at a representative range of $(T,\mu)$ values, also indicated by the yellow circles in Fig.~\ref{fig:phase}.}
\end{figure*}

 Figure \ref{fig:dists} shows note length probabilities $P_t$ calculated at 30 $(T,\mu)$ points, indicated by yellow points on the phase diagram in Fig.~\ref{fig:phase}. The points are chosen so as to illustrate the full range of typical behavior. In general, the panels in Fig.~\ref{fig:dists} are arranged with temperature increasing to the right and $\mu$ becoming more negative towards the bottom. The lowest temperature points in the left column of Fig.~\ref{fig:dists} all show nearly uniformly repeating notes of the same length, here denoted as eighth notes. Going along any row or down any column in Fig.~\ref{fig:dists}, an evolution of the $P_t$ values can be discerned. As $T$ increases, the number of different note lengths represented increases. At higher $\mu$ (less negative), the additional note lengths tend to be shorter than the original eighth notes, as a higher note concentration is favored. At lower $\mu$ (more negative), the additional note lengths tend to be longer than the original eighth notes, as lower note concentration is favored. In all cases, the additional note lengths are most likely to be related to the original eighth notes by factors of two, as expected for the common time metric hierarchy.

 The results shown in Fig.~\ref{fig:dists} put constraints on the types of note length distributions we expect to see in music. In all cases, the note length distributions are dominated by one or two note lengths. This corresponds to the top level of the metric hierarchy with $p_l \approx 1$, and at most one other level of the hierarchy with significant $p_l$. This is illustrated by the $p_l$ values shown in Fig.~\ref{fig:tsweep}. Even at the highest $T$ (Fig.~\ref{fig:tsweep}(d)) there is only a slightly visible contribution from the third level of the hierarchy, and the values of $P_t$ are still dominated by eighth and sixteenth notes. Further increase of $T$ does not lead to more significant levels of the hierarchy; it instead leads to the transition to the disordered phase.

 We can quantify, out of the space of all possible sets of $p_l$, which sets are similar to solutions of the model. Given a particular set of note onset probabilities $\tilde{p}_l$, its similarity to a result from the model $p_l(T,\mu)$ can be quantified as
\begin{equation}
\sigma(T,\mu) = \left[ \sum_{l=0}^{L-1} \left( p_l(T,\mu) - \tilde{p}_l \right)^2 \right]^{1/2}.
\end{equation}

We can then find the minimum value $\sigma_0 = \sigma(T_0,\mu_0)$ over a range of computed model results, in this case the range of $\mu$ and $T$ shown in the phase diagram in Fig.~\ref{fig:phase}. (See Supplementary Material for further details.) If we randomly sample over the space of possible $\tilde{p}_l$ we calculate a mean value $\bar{\sigma}_0 = 0.72$ with standard deviation 0.15. This indicates that with the $\tilde{p}_l$ each selected from a uniform random distribution, the set of $\tilde{p}_l$ is extremely unlikely to be similar to a result of the model, in which case $\sigma_0 \approx 0$. We can compare the model results to a restricted space of $\tilde{p}_l$, where we randomly sample only $\tilde{p}_l$ with the type of hierarchical structure shown in Fig.~\ref{fig:meter}. In this case, we calculate $\bar{\sigma}_0 = 0.23$ with standard deviation 0.11. By presupposing the hierarchical structure we find smaller mean error $\bar{\sigma}_0$ but still more than two standard deviations from $\sigma_0=0$. This shows that only certain hierarchical note onset probabilities are predicted by the model, which we will compare to results from musical examples below.

For completeness, here we describe two orderings that appear with different character in some high-temperature regions of the phase diagram. Once case is the narrow dark red strip that occurs at the highest $T$ before the transition to disorder, and which fades to black as $\mu$ becomes more negative. This phase is characterized by all $p_l \ll 1$ with alternating amplitude. The resulting note length probabilities are quite similar to those of the disordered case, though with an alternating deviation from the disordered exponential distribution. The other case is the bistable solution that emerges at high $T$, such as the green dots within the blue region at high $T$ in Fig.~\ref{fig:phase}. This solution is also similar to the disordered phase, but with increased spacing between nonzero $p_l$. For example, the high-$T$ green points in the blue region have $p_0 = p_4 \sim 0.5$ and other $p_l \approx 0$. This yields a note length distribution with prominent $P_4, P_8$, $P_{12}$, etc. decaying exponentially, and other $P_t \approx 0$. It is unclear whether these two cases are seen commonly in music. As we will see below, they are not clearly represented in the examples studied here. Given their proximity to the disordered phase, it is possible that such rhythms appear in music that is less regularly ordered.

In the $L=2$ and $L=8$ results presented above, we have seen similar behavior but with the larger $L$ allowing more levels of the hierarchy to emerge. As such, we can expect that further increasing $L$ to a larger power of two will yield qualitatively similar results. When $L$ is divisible by factors other than two, then new orderings become possible. For example with $L=6$, levels of the hierarchy may arise from divisions by two or by three. This case is presented in detail in the Supplementary Material. In general, the results show substantial bistability between solutions involving divisions by two or by three. For example, at the same values of $\mu$ and $T$, one might find $p_l\approx1$ at every second or every third value of $l$, with the other $p_l$ lower.

In the $L=6$ case, however, we observe a tendency towards division by two. For example, when the strong beat has $p_l\approx 1$ at every sixth $l$, the next level of the hierarchy occurs by a division of two and not three. As shown in the Supplementary Material, this is to be expected from a free energy argument. A single division of the strong beats by two creates many rhythmic triples with the strong beats, whereas a single division by three requires two new notes to create just one rhythmic triple. Qualitatively, this matches the meters typically used in music. The divisions by two represent the meter of ``common time'' or 4/4 time. Other less common meters, such as 3/4 (three quarter notes per measure) or 6/8 (six eighth notes per measure) permit division of the beat by three. These meters do not require that the primary beats are divided by three, for example in 3/4 time if the primary beats are quarter notes. But it is possible that the primary beat in 3/4 time is a dotted half note, which would then be divided into three weaker quarter note beats. We can expect that $L$ divisible by larger primes (5, 7 etc.) will in principle allow meters with divisions of those larger numbers, but will be increasingly disfavored compared to meters with smaller divisions. Musically, meters with such larger divisions (e.g. 5/4) do occur, but it is extremely uncommon for the primary beat to be so finely divided.

\subsection{Comparison: Bach Cello Suites}

To illustrate how our model captures the rhythmic patterns that occur in music, we compare the model results to rhythms in the six solo cello suites by Johann Sebastian Bach. These pieces were selected for several reasons. First, the six suites are each comprised of six movements with the same names: I. Prelude, II. Allemande, III. Courante, IV. Sarabande, V. Menuet/Bourree/Gavotte 1/2, and VI. Gigue, with the exception of movement V. (The name of Movement V varies, and is also divided into two parts that we will analyze separately below.) Following the initial Prelude, the name of each movement refers to a particular baroque dance style.~\cite{Marckx1998,Harlow1993} This structure of the Suites yields a dataset of 42 movements (or parts of movements) that permits comparison of the same movement across 6 suites in a similar style, and between the 7 movements (or parts) in a single suite in varying styles. It is also convenient that these pieces are written for a single voice (solo cello) that typically sounds only one note at a time. Furthermore, each movement is relatively short, with a consistent rhythmic character throughout each movement. (The one exception to this is the Prelude from Suite 5 which begins with a slower section (Adagio), then transitions to a distinct faster section (Allegro). For simplicity we will only consider the Allegro section below.) These factors simplify the analysis of the rhythm. 

\begin{figure*}
\centering
\includegraphics[width=1.0\textwidth]{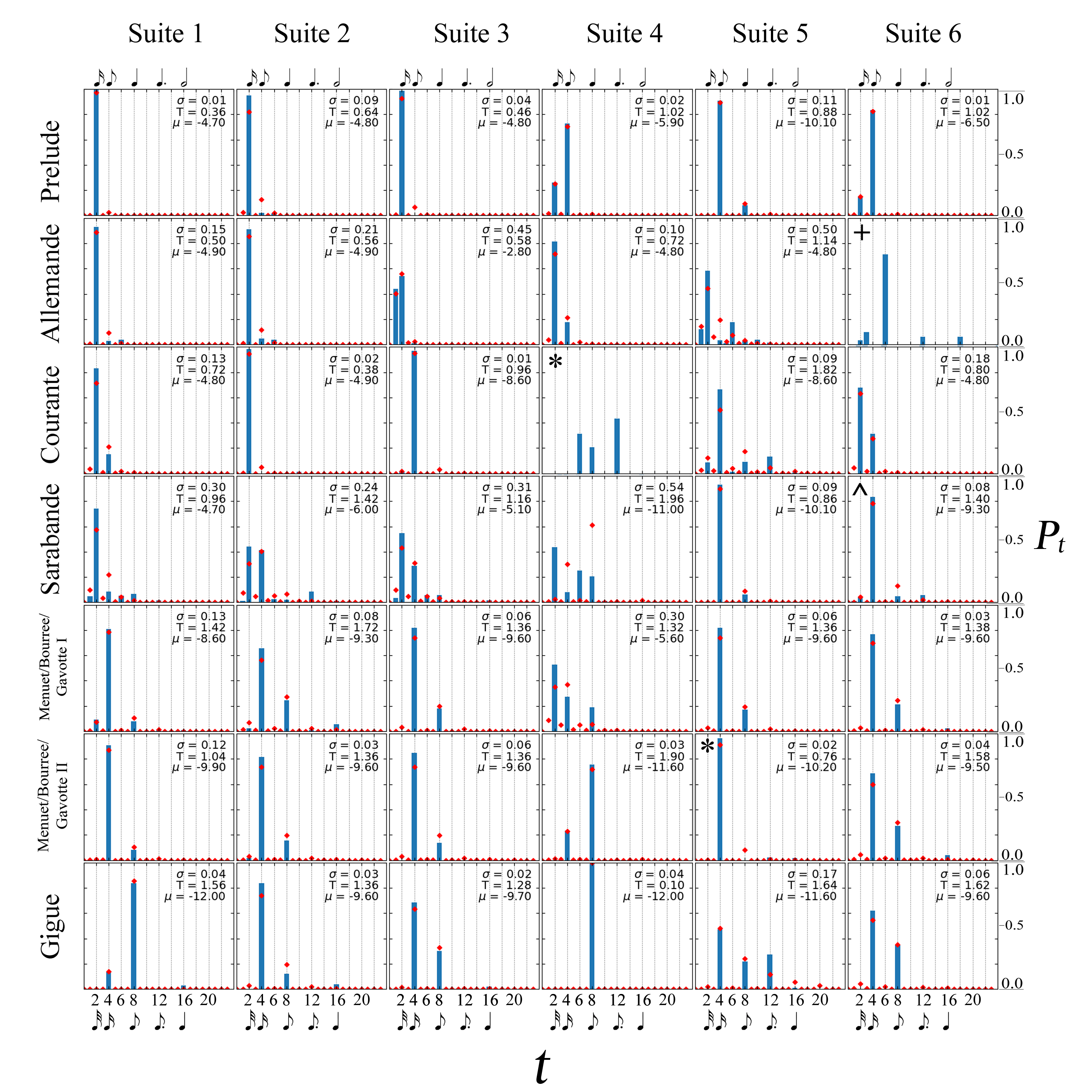}
\caption{\label{fig:cello}Probabilities $P_t$ of note lengths $t$ for each movement of the Bach's six cello suites (BWV 1007–1012). Red, purple, and green lines indicate commonly occurring note lengths. Note lengths are as indicated along the top or bottom horizontal axes, except for panels marked `$\ast$', `+', and `$\wedge$', which have minimum note lengths of a triplet 32$^\mathrm{nd}$ note, a triplet 128$^\mathrm{th}$ note, and a 16$^\mathrm{th}$ note, respectively. The Prelude from Suite 5 and all Gigues use the note length labeled on the bottom axis because they are notated using eighth note beats instead of quarters, all other plots refer to note lengths at the top axis. In all cases, the maximum note length shown is 24 times the minimum. The Prelude in Suite 5 excludes the opening Adagio section.}
\end{figure*}

We obtained the note length probabilities $P_t$ shown in Figure~\ref{fig:cello} by counting the spacing between note onsets (changes in pitch or re-articulations of the same pitch), as described in the Methods section below. We can immediately see from Fig.~\ref{fig:cello} that the model's prediction of one or two dominant note lengths is generally accurate. The relative relationships of these lengths are also consistent with our model, with the lengths of most significant $P_t$ being related by factors of $2$ or $1/2$. 

In order to compare the observed note length distributions to the results of the $L=8$ model, we compute the minimum error $\sigma_0(T_0,\mu_0)$ as described above, for note onset probabilities $\tilde{p}_l$ extracted from the music. The details are given in the supplementary material. With this method, we obtain the minimum error $\sigma_0$ and the values of $T_0$ and $\mu_0$ at which the minimum occurs for 40 out of 42 movements, with the Allemande from Suite 6 and Courante from Suite 4 excluded due to presence of triplets that confound this $L=8$ analysis. The resulting values are listed in the corresponding panels of Fig.~\ref{fig:cello}. Averaging over all 40 analyzed movements, we obtain a mean error $\bar{\sigma}_0 = 0.12$ with standard deviation 0.13. This agreement is substantially better than in the randomly sampled hierarchies discussed above. As shown in Fig. S4 in the supplementary material, the values of $\sigma_0$ from the Bach Cello Suites are clustered near $\sigma_0=0$, unlike the cases of randomly sampled $\tilde{p}_l$.

From the model $p_l$ at the best fit values of $T_0$ and $\mu_0$, we can calculate the predicted note length distribution for each movement, shown as red dots in Fig.~\ref{fig:cello}. In many cases, we see excellent agreement. Note that the predicted note length distribution is not obtained by minimizing the deviation from the observed note length distribution $P_t$. It is obtained instead by a stricter condition: minimization of the deviation of the pattern of note onset probabilities $p_l$. This includes not just the frequency of note length occurrences, but also where in the rhythm those notes occur.

Looking closer at the data in Fig.~\ref{fig:cello}, we can compare the observed $P_t$ in different movements to the predicted distribution. We see that the model results can capture the rhythmic patterns seen in many of the Cello Suites. For example, we see that the first three Preludes (top row of Fig.~\ref{fig:cello}) correspond to low temperatures, with one note length overwhelmingly dominant. The deviation between the observed and predicted note length distributions in the Prelude from Suite 2 shows how the model predicts only a restricted set of possible distributions. In this case, the movement consists predominantly of sixteenth notes, resulting in higher values of every second $p_l$. Given these values, the model predicts a certain frequency of eighth notes, shown by the red dot at $P_4$. While eighth notes are, in fact, the second most common note length in the movement, they are less frequent than the model predicts. In the Preludes from Suites 4, 5 and 6, there is a second prominent length at half or double the most common length, indicating somewhat higher $T$. The Preludes in Suites 4 and 6 are mostly eighth notes with some sixteenth notes, whereas the Prelude in Suite 5 (Allegro section) is mainly sixteenth notes with some eighth notes. In these cases, we see that the frequency of both note lengths are captured quite well by the model.

In contrast to the regular rhythms of the Preludes, the Sarabandes in Suites 1-4 have a greater number of significant note lengths. The Sarabandes in the Cello Suites are slower movements, often with a freer rhythm~\cite{Marckx1998}. This often corresponds to a relatively higher temperature. Given the greater complexity of these distributions, we are less likely to find precise quantitative agreement with the predicted distribution. Nonetheless, the predicted $P_t$ in Suites 1-3 do capture the general trends of the several most prominent note lengths. The Sarabande from Suite 4 is a notable exception, in that the model here fails to predict the correct note length distribution.

Other movements more closely match the intermediate case where one note length is dominant, with small but noticeable occurrences of notes at half and double the primary length. For example, Menuet I from Suite 1 is comprised mainly of eighth notes with about equal numbers of sixteenth notes and quarter notes. The $P_t$ predicted by the model accurately captures the frequency of eighth notes, and also reproduces the presence of sixteenth and quarter notes. Even the small but noticeable number of dotted quarter notes is reproduced here. As another example, the Menuet II from Suite 2 is mainly eighth notes, with some quarter notes, and a small but noticeable number of sixteenth notes, all of which are well captured by the model.

\section{Discussion}

The comparison above between the note length probabilities $P_t$ calculated from the model and observed in Bach's Cello Suites (Fig.~\ref{fig:cello}) demonstrates qualitative agreement. The general characteristics of the metric hierarchy are seen in both, and typically one or two dominant note lengths are seen in both. In many cases, the agreement is quantitatively good, with minimum error $\sigma_0 < 0.1$. There are, however, some discrepancies between the model and the data. Of course, this is to be expected to some degree. One limiting factor is that the rhythms in each movement provide a finite sample of note lengths. Though there may be hundreds of notes in a given movement, the rhythmic motives tend to repeat approximately or exactly, meaning that the effective sample size of independent note lengths is much smaller than the total number of notes. Other discrepancies between the model and the data can be attributed to assumptions made in the model.

One simplifying assumption above is that we have compared the Cello Suites to the $L=8$ model only. At first sight, it may seem surprising that we see good agreement between the $L=8$ model and the Bach Cello Suites when we consider that about half of the movements are written in meters that involve division by three. However, this mystery is resolved by the predicted and observed fact that the note lengths are governed by typically one or two levels of the metric hierarchy. In the case of nearly all movements in the Cello Suites, the levels of the hierarchy governing the rhythm lie below any divisions of three in the meter. In music theory, the distinction is made between ``simple triple'' meters in which the primary beats are organized with a factor of three, but the beats themselves are divided by twos, and ``compound'' meters in which the primary beats are divided by a factor of three. For example, the Courante in Suite 1 is written in 3/4 time (a simple triple meter), meaning that each measure is divided into three quarter notes, and those quarter notes are subdivided into shorter notes by factors of two. The note lengths in that movement are almost entirely sixteenth notes and eighth notes, in the levels of the metric hierarchy divided by twos. In such cases, the division by three in the meter doesn't significantly affect the distribution of note lengths, but will affect other elements of the music such as the melody, harmony, and pattern of stresses of notes. If the rhythm were drawn from higher levels of the metric hierarchy, then we might see mostly dotted half notes (three quarter note spans) with some quarter notes. Such a rhythm would not be predicted by the $L=8$ model shown here. Perhaps the only place where division by three clearly affects the rhythms seen here is in the Gavotte II in Suite 5. This movement (though actually written in 2/4 time) is comprised mainly of triplet eighth notes effectively yielding a compound meter. The next shortest note lengths in the movement are quarter notes, three times longer. This factor of three is seen in Fig.~\ref{fig:cello} with the main bar at note length 4, and the next small bar at length 12. The other cases that include divisions of three are the Courante in Suite 4 and the Allemande in Suite 6. In these cases, the overall meter contains only divisions of 2, but isolated triplets appear in several places. This gives rise to an extra note length in the $P_t$ plot smaller than the others. In the Courante in Suite 4, there are a significant number of triplets, yielding a relatively large bar at the leftmost position in Fig.~\ref{fig:cello}, with relatively fewer triplets seen in the Allemande from Suite 6. In both cases, if we ignore that shortest-length, leftmost bar, the rest of the $P_t$ agree well with the typical results of the model. These isolated triplets represent a local change in the meter, which cannot be captured in the mean field model here with infinite range interaction. 

An interesting result to emerge from the model presented here is the presence of syncopated rhythms. Syncopated rhythms are rhythms that deviate from the most probable patterns suggested by the metric hierarchy. Due to the balance between rhythmicity and variety, the model does, in fact, predict the occurrence of syncopated rhythms. This occurs when a time bin with high $p_k$ does not contain an event, and is preceded or followed by a bin with lower $p_k$ that does contain an event. When the occupied lower $p_k$ bin precedes the unoccupied higher $p_k$ bin, the effect of syncopation is stronger. The opposite order produces a weaker syncopation such as a dotted-eighth-sixteenth note pattern. At higher $T$, where variety is preferred relative to predictability, the highest $p_k$ are significantly less than unity and hence syncopation is more common. 

An example where syncopation occurs in line with the model is the Sarabande in Suite 3, with a noticeable bar at $P_6$, a dotted eighth note. Other movements use syncopated rhythms almost exclusively. For example, the Gigue in Suite 5 has weakly syncopated (dotted-eighth-sixteenth) rhythms in almost every measure. The $P_t$ for this movement in Fig.~\ref{fig:cello} shows a value of $P_{12}$ (dotted eighth note) larger than would be predicted by the model here. Syncopated rhythms like these provide a way to study the balance between predictability and surprise~\cite{Huron2008, London2012}, and the role of preference for the perception of particular rhythmic priors~\cite{Kaplan2022,Jacoby2024}.

This work has shown how familiar musical meter can emerge from a simple model based on a few foundational assumptions -- that humans have a preference for perceiving patterns of repeated events, that humans also have a preference for variation and complexity in those patterns, and that rhythms result from a balance between those preferences. Despite the simplicity of these assumptions, the ordered patterns of events that emerge bear striking resemblance to rhythms observed in music. 

It is well-known that the structure of musical meter is hierarchical, but not understood whether that structure is innately built into the human perception of rhythm. Our results show that hierarchical structure need not be explicitly represented in a perceptual model, but can emerge from the simple preference for 1:1 repetition of time intervals.

The results here provide a quantitative link between the hierarchical structure of meter, and the rhythms produced based on those meters. Previous work has described the qualitative tendencies of how meter governs rhythm, for example the increased likelihood of note onsets at stronger time points~\cite{lerdahl1996generative}. Some work has attempted to quantify these probabilities empirically~\cite{Hutchinson1987, Palmer1990, Temperley1999,Temperley2007}. In contrast, our results provide a bottom-up calculation of these probabilities that shows qualitative agreement with the dataset presented here (the Bach Cello Suites).

The view of rhythm and meter as an ordered phase of sound opens new avenues for understanding and creating music. Our discussion of the Bach Cello Suites above demonstrates how this, or related, models can be used as tools for quantitative musicology. Qualitative statements about the rhythmic character of a piece or a passage can be put on firmer ground by comparing to predicted rhythms at a particular temperature or chemical potential. In the future, more sophisticated statistical techniques, such as Bayesian analysis, could be used to further quantify how observed rhythms correspond to predicted meters. In the other direction, these results provide a mechanism for generating new rhythms, for algorithmic music composition or live performance.

\section{Methods}

\subsection{Normalization of total rhythmicity}

In the development of Eq.~\ref{eq:Rtot} for $R_{tot}$ above, we set the prefactor $R_0 = r_0$ for $k-j \leq N$ and $R_0=0$ for $k-j>N$.  To motivate the choice of $r_0 = 1/(N\tilde{B}^2)$, we consider an example where every $m^\text{th}$ $B_i = 1$ and all other $B_i=0$. Since the first sum is over an infinite range, we will consider $\tilde{R}_{tot}$ as the $R_{tot}$ per time bin. For simplicity, we will also assume that $N$ is a multiple of $m$. There is then a fraction $1/m$ of nonzero $B_j$. At those $j$, there are $N/m$ nonzero terms of $B_{j-s}B_{j+s}$ in the second sum. This yields $\tilde{R}_{tot} = r_0 N/m^2$. If $r_0$ were constant, this would imply a strong dependence of $\tilde{R}_{tot}$ on $m$, with the fastest possible rhythm always being the most rhythmic. As mentioned above, however, we might expect a lower $R_0$ for a rhythm that is more dense in time (smaller $m$). To avoid the need for a complicated psychoacoustic model with multiple timescales, if we take $r_0 = m^2/N$ in this example, then we obtain the particularly simple $\tilde{R}_{tot} = 1$, meaning that any pattern of regularly repeating beats with period of anywhere from 1 to $N$ sites would have the same $\tilde{R}_{tot}$. The decision to normalize $r_0$ by $N$ is not required since $N$ is a constant, but will simplify expressions. 

The choice of $r_0 = m^2/N$ in the above example motivates us to more generally choose $r_0 = 1/(N\tilde{B}^{2})$, where $\tilde{B}$ is the mean of the $\{B_i\}$ -- the average occupancy of the time bins. This would yield $r_0 = m^2/N$ in the example above, and more generally yields a partial scale invariance where $\tilde{R}_{tot}$ is insensitive to scaling a periodic rhythm by a constant factor, so long as the period is less than $N$. 

\subsection{Calculation of Landau free energy}

Solving Equation \ref{eq:nmdiff} for $h$ we obtain

\begin{equation}
\label{eq:h}
\left. \frac{\partial F}{\partial m} \right\rvert_{\mu,T} =h = -2\frac{m}{n} + \frac{1}{2} T\log{\left[ \frac{1-(n-m)}{n-m}\frac{n+m}{1-(n+m)} \right]}.
\end{equation}

Additionally, we can rearrange Equation \ref{eq:nmsum} to obtain an equation for $\mu$:

\begin{equation}
\mu = -3 - \left( \frac{m}{n} \right)^2 + \frac{1}{2} T\log{\left[ \left(\frac{1-(n-m)}{n-m}\right) \left(\frac{1-(n+m)}{n+m}\right) \right]}. 
\label{eq:mu}
\end{equation}

We can use this equation to obtain contours of constant $\mu$. This is essentially the same as contours of Equation \ref{eq:nmsum} plotted to graphically solve for $n$ and $m$.

We can then obtain the free energy $\mathcal{F} = F(\mu, T, m)$ at fixed $T$ and $\mu$ by first numerically converting $\frac{\partial F}{\partial m} (n,T,m)$ to $\frac{\partial F}{\partial m} (\mu,T,m)$ using $\mu(m,T,n)$ given in Eq.~\ref{eq:mu}, then numerically integrating $\frac{\partial F}{\partial m} (\mu,T,m')$ from $m'=-1/2$ to $m$. This yields the Landau free energy $\mathcal{F}(m)$ at a particular $\mu$ and $T$.

\subsection{Solving the L-site model}
We find solutions to Eq.~\ref{eq:dR with p} and Eq.~\ref{eq:p} for $L>2$ using an iterative method. First, the $L$ values of $p_l$ are initialized to some values $p_l^0$. For example, $p_l^0$ can be chosen as uniformly random numbers $0 < p_l^0 \leq 1$. In general we find that larger $L$ can result in larger multi-stability of solutions. As such, the initial values $p_l^0$ can also be chosen to bias towards particular solutions, if desired. These values are plugged into the right-hand side of Eq.~\ref{eq:p}, yielding values $p'_l$. New values are then obtained as $p_l^1 = \alpha p'_l + (1-\alpha) p_l^0$. $0<\alpha\leq1$ is a weighting factor that controls how quickly the iteration converges. $\alpha<1$ can prevent the solutions from oscillating instead of converging. The $p_l^1$ are then plugged into the right hand side of Eq.~\ref{eq:p}, and the process is iterated. Repeated iteration results in convergence to self-consistent values of $p_l = p_l^\infty$ at a stable solution. In practice, the iteration is halted when the mean squared change between $p_l^n$ and $p_l^{n-1}$ falls below a small threshold.

\subsection{Extraction of note lengths from midi files}
Note lengths were extracted from midi recordings of all movements of Bach's Cello Suites. Files were first cleaned to remove ornamentations such as trills and grace notes, and were manually checked against the written score. ``Note lengths'' were then obtained as the time interval between successive note onsets. These note lengths differ slightly from the distribution of actual written note lengths, in that rests are included in the inter-onset time interval. In Fig.~\ref{fig:cello}, the Prelude from Suite 5 and all Gigues were set to have a minimum note length of a $64^\mathrm{th}$ note (bottom axis), while the others are set to have a $32^\mathrm{nd}$ note as the minimum division (top axis), except as otherwise indicated. Note lengths are shown up to 24 times the minimum length, with the maximum length typically corresponding to a dotted half note (top axis), or dotted quarter note (bottom axis), unless otherwise indicated. Note lengths beyond the plotted range do occur, but very rarely, and almost always at the end of a movement or end of a major section. As such, they serve more to end the rhythmic passage than to take part in it. There are several movements that required slight variations on the analysis method and are indicated by `$\ast$', `+' and `$\wedge$' in Fig.~\ref{fig:cello}. The two movements marked with `$\ast$' include triplets (three-note patterns that fill a time span filled by two notes in the normal meter.) These movements have been analyzed with a smaller time bin to accommodate these divisions by either 2 or 3. The movements marked `+' and `$\wedge$' are written with particularly short and long note lengths, respectively. As such the movement marked `$\wedge$' has a minimum note length of a sixteenth note and a maximum note length of a whole note. The movement marked `$+$' contains both $64^\mathrm{th}$ notes, and $64^\mathrm{th}$ note triplets, so uses a particularly small time bin to accommodate both of these fine time divisions.

\subsection{Generation of audio examples}

Given a set of note onset probabilities $p_l$ from a solution to the $L$ site model, we generate a rhythm by sampling from these probabilities. At time bin $k$, a note onset is chosen or not according to the probability $p_k$, where the $L$ independent values of $p_l$ repeat periodically. These note onset times are used to generate a midi file, with the note volume (or velocity) modulated by the value of $p_l$ to include accents that follow the calculated meter. These midi files are then rendered to a .wav file using a xylophone audio sample.

\section*{Data Availability Statement}
Software used for solving the $L$-site mean field model, comparing model results to musical examples, and plotting figures is available at 10.5281/zenodo.19473679.

\section*{Author Contributions}
St. Clair: Software, Investigation, Visualization, Writing - Original Draft, Berezovsky: Conceptualization, Methodology, Software, Formal Analysis, Supervision, Writing - Review \& Editing, Funding Acquisition.

\begin{acknowledgements}
The authors acknowledge the support from the Case Western Research University Expanding Horizons Initiative in the College of Arts and Sciences and the Jack, Joseph and Morton Mandel Foundation through an Experimental Humanities (EH1-S) award.
\end{acknowledgements}

\bibliography{rhythm}

\end{document}